\documentclass[apj, iop]{emulateapj-rtx4}
\usepackage{natbib}

\shorttitle{Asymmetric Ejecta Distribution in SN\,1006}
\shortauthors{Uchida et al.}

\begin{document}

\title{Asymmetric Ejecta Distribution in SN\,1006}

\author{Hiroyuki~Uchida\altaffilmark{1}, Hiroya~Yamaguchi\altaffilmark{2}, and Katsuji~Koyama\altaffilmark{1,3}}
\email{uchida@cr.scphys.kyoto-u.ac.jp}

\altaffiltext{1}{Department of Physics, Graduate School of Science, Kyoto University, Kitashirakawa Oiwake-cho, Sakyo-ku, Kyoto 606-8502, Japan}
\altaffiltext{2}{Harvard-Smithsonian Center for Astrophysics, 60 Garden St., Cambridge, MA 02138, USA}
\altaffiltext{3}{Department of Earth and Space Science, Graduate School of Science, Osaka University, 1-1 Machikaneyama, Toyonaka, Osaka 560-0043, Japan}

\begin{abstract}
We present the results from deep X-ray observations ($\sim400$\,ks in total)
of SN\,1006 by the X-ray astronomy satellite Suzaku.  
The thermal spectrum from the entire supernova remnant (SNR) exhibits prominent
emission lines of O, Ne, Mg, Si, S, Ar, Ca, and Fe.
The observed abundance pattern in the ejecta components is in good 
agreement with that predicted by a standard model of Type Ia supernovae (SNe).
The spatially resolved analysis reveals that the distribution of
the O-burning and incomplete Si-burning products (Si, S, and Ar) is 
asymmetric, while that of the C-burning 
products (O, Ne, and Mg) is relatively uniform in the SNR interior. 
The peak position of the former is clearly shifted by $5\arcmin$ ($\sim$3.2\,pc at 
a distance of 2.2\,kpc) to the southeast from the SNR's geometric center.
Using the SNR age of $\sim$1000\,yr, we constrain the velocity 
asymmetry (in projection) of ejecta to be $\sim$3100\,km\,s$^{-1}$.    
The abundance of Fe is also significantly higher in the southeast region than 
in the northwest region. 
Given that the non-uniformity is observed only among the heavier elements (Si through Fe), we argue that SN\,1006 originates from an asymmetric explosion, as is expected from recent multi-dimensional simulations of Type Ia SNe, although we cannot eliminate the possibility that an inhomogeneous ambient medium  induced the apparent non-uniformity.
Possible evidence for the  Cr K-shell line and line broadening in the Fe K-shell emission is also found.

\end{abstract}


\section{Introduction}
Despite many efforts in the last decades, the explosion mechanism of Type Ia supernovae 
(SNe) is still unclear. 
It is widely known that Type Ia SNe show significant diversity in the optical spectra
and the light curves \citep[e.g.,][]{Phillips1999, Benetti2005}. 
Theoretically, the diversity has been interpreted as a consequence of spherically 
asymmetric explosions (e.g., \citealt{Kasen2009}; also see 
\citealt{Mazzali2007}). 
\citet{Maeda2010a} systematically studied Type Ia SNe
and attributed the observed spectral diversity to random
viewing angles in almost identically symmetric explosions.
Besides them, multi-dimensional simulations have suggested 
that thermonuclear ignition in Type Ia progenitors is offset from the center 
\citep[e.g.,][]{Woosley2004, Kuhlen2006, Roepke2007}, which may result in 
a non-uniform distribution of the nucleosynthesis products.

Young supernova remnants (SNRs) in our Galaxy are ideal sites to investigate 
the abundance and distribution of SN ejecta in detail, because they are spatially 
well resolved, unlike extragalactic SNe. 
SN\,1006, one of the prototypical Type Ia SNRs, is particularly appealing for 
such investigation owing to its proximity \citep[2.2\,kpc: ][]{Winkler2003} and 
moderate angular size ($30\arcmin$ in diameter). 
It is located at a high Galactic latitude ($b = 14.6$) and hence has small foreground 
extinction. 
SN\,1006 is therefore a suitable object to investigate the spatial information of 
low-$Z$ elements, such as O, Ne, and Mg, for which K-shell emissions are observed in 
soft X-rays. 
Its low and uniform ambient density \citep{Dubner2002} also enables us to study its nearly  
pure distribution of ejecta, with no significant modification owing to the swept-up 
interstellar medium (ISM). 
Nonetheless, our knowledge of  ejecta has been limited because previous studies 
on SN\,1006 have been mostly on the northeast (NE) and southwest (SW) rims, where the 
cosmic-ray acceleration (non-thermal X-ray) is dominant \citep[e.g.,][]{Koyama1995, 
Bamba2003, Cassam2008}. 
 
Although Type Ia SNe yield a large amount of Fe \citep[e.g.,][]{Iwamoto1999}, 
detection of any signature of  Fe ejecta from SN\,1006 is hampered.
One reason is that the thermal X-ray radiation from ejecta is faint compared 
to the bright non-thermal emissions, mainly from the cosmic-ray-accelerating rims. 
Furthermore, considerable fractions of  ejecta may still not be shocked
to emit X-rays.  
In fact, blue- and red-shifted absorption lines of \ion{Fe}{2} were detected in the 
UV spectra of the background stars, suggesting that un-shocked Fe ejecta freely 
expands in the SNR's interior \citep{Wu1993, Hamilton1997, Winkler2005}.
However, the inferred  amount of cool Fe ($< 0.2M_\odot$) was much less than that 
of the theoretical prediction for a typical Type Ia SN ($\sim 0.7 M_\odot$). 
Thus, a large amount of shocked Fe may exist in the SNR.
Possible evidence for the shocked Fe ejecta (i.e., X-ray line emission) in SN\,1006 
was first reported from the \textit{BeppoSAX} observation \citep{Vink2000}. 
\citet{Yamaguchi2008} found clearer evidence for the Fe K-shell line at $6.43 
(\pm0.02)$\,keV using \textit{Suzaku}. 
The mean ionization age of Fe corresponding to this
centroid energy was found to be much lower than those of the
other lighter elements (e.g., Si, S), suggesting that the
Fe-rich core was heated by the reverse shock more recently.

\citet{Yamaguchi2008} also revealed that the Fe K-shell lines are the brightest in the southeast (SE) quadrant, 
which may imply an asymmetric distribution of Fe-rich ejecta. 
However, the detailed spectroscopy was limited within the SE region, and 
spatial distributions of the other lighter elements were 
not well investigated.  

In this paper, we report a comprehensive study on the shocked ejecta in SN\,1006
using the deep exposure ($>400$\,ks) data of \textit{Suzaku}.
Owing to the high sensitivity and good energy resolution, 
we are able to reveal detailed distribution of  ejecta. 
Throughout this paper, the distance to SN\,1006 is assumed to be 2.2\,kpc 
\citep{Winkler2003},  and errors are quoted at a 90\% confidence level unless 
otherwise noted.

\begin{table*}[!t]
\caption{Observation logs.}\label{tab:obs}
\begin{center}
\begin{tabular}{llcccc}
\tableline	   						    		      \tableline	   						    		      			  						
& Name & Obs. ID  & Obs. Date & (R.A., Dec.) $_{J2000}$ & Exposure\\
\tableline	   						    		      			  						
Source  & NE    & 100019020   & 2005-Sep-09 &  (225.9608, -41.7805)  & 22\,ks\\ 
 & SW1  & 100019030   & 2005-Sep-10 &  (225.5010, -42.0706)  & 31\,ks\\ 
 & SW2  & 100019050   & 2006-Jan-26 &  (225.4998, -42.0701)  & 31\,ks\\ 
 & SE  & 500016010   & 2006-Jan-30 &  (225.8686, -42.0508)  & 52\,ks\\ 
 & NW  & 500017010   & 2006-Jan-31 &  (225.6397, -41.7993)  & 53\,ks\\ 
 & Center  & 502046010   & 2008-Feb-25 &  (225.7268, -41.9424)  & 212\,ks\\ 
 \tableline
Background & NE\_BGD & 100019010   & 2005-Sep-04 &  (226.7036, -41.3998)  & 45\,ks\\ 
 & SW\_BGD1 & 100019040   & 2005-Sep-11 &  (224.6550, -42.4005)  & 32\,ks\\ 
 & SW\_BGD2 & 100019060   & 2006-Jan-26 &  (224.6468, -42.4025)  & 28\,ks\\ 
\tableline
\end{tabular}
\end{center}
\end{table*}

\section{Observations and Data Reduction}
\begin{figure}[t]
\begin{center}
\includegraphics[width=90mm]{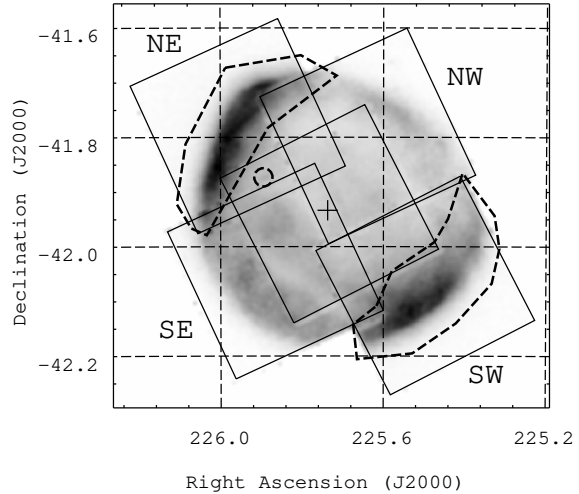}
\caption{Vignetting-corrected XIS image of SN\,1006 in the 0.4--5.0\,keV band. The FoVs of XISs are 
shown with the solid squares. The cross mark indicates the geometric center of SN\,1006. 
The dashed line regions are  bright non-thermal rims with a background point 
source \citep[QSO 1504-4152;][]{Winkler1997}, which are excluded for the thermal spectrum analysis. 
}
\label{fig:fov}
\end{center}
\end{figure}

We performed several pointing observations of SN\,1006 with the X-ray Imaging 
Spectrometer \citep[XIS;][]{Koyama2007} on board \textit{Suzaku} 
\citep{Mitsuda2007}. 
The series of the observations covers almost the entire region of the SNR (Figure~\ref{fig:fov}). 
The observations of the four quadrant regions were made during the performance 
verification (PV) phase by the \textit{Suzaku} Science Working Group (SWG). 
To compensate for the relatively low sensitivity near the edge of each field of view 
(FoV) and to study more about the ejecta in the SNR interior, we performed a deeper 
observation aiming at the SNR's center during 
the Announcement of Opportunity cycle 2 (AO2) phase. 
Detailed information about the observations made is summarized in Table~\ref{tab:obs}.
We used XIS1 (back-illuminated CCD; BI CCD), and XIS0, 2, and 3 (front-illuminated 
CCDs; FI CCDs) for the PV-phase data.
However, for the AO2, XIS2 data were not available 
owing to possible damage by a micrometeorite on November 9, 2006. 
We used revision 2.4 of the cleaned event data and combined the $3\times3$ and 
$5\times5$ pixel events. 
The calibration database (CALDB) updated in September 2011 was used for  data 
reprocessing. 
We performed data reduction with  version 6.11 of the HEAsoft tools (version 18 
of the \textit{Suzaku} software).

\begin{figure*}[th]
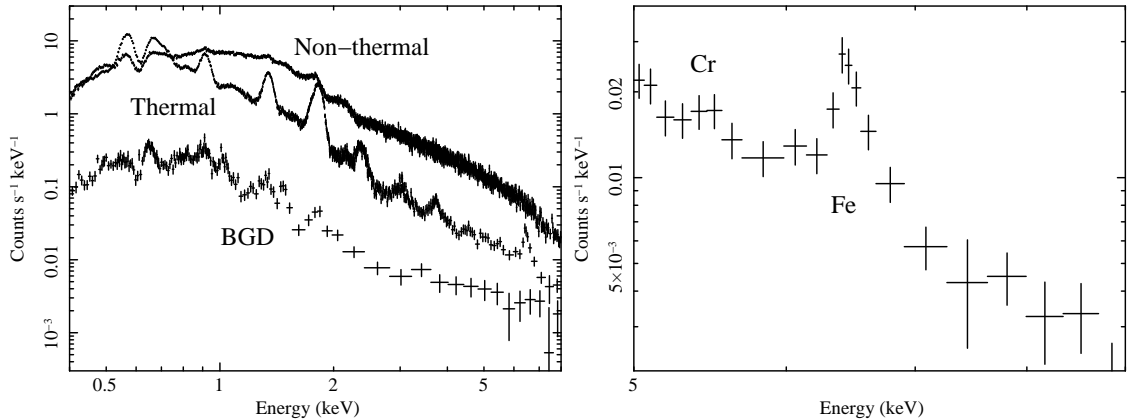

\begin{center}
\includegraphics[width=55mm,angle=-90]{figure2a.eps}
\includegraphics[width=55mm,angle=-90]{figure2b.eps}
\caption{\textit{Left}: FI spectra of the thermal and non-thermal regions in 
SN\,1006, where the NXB and CXB are subtracted. The spectrum of the BGD region is 
also shown. \textit{Right}: The magnified spectrum of the thermal region in 
5.0--10.0\,keV.}
\label{fig:thermal}
\end{center}
\end{figure*}

\section{Analysis and Results}
We employed the \textit{xisrmfgen} XIS response generator and the \textit{xissimarfgen} ray-tracing-based ancillary response file generator to generate the 
redistribution matrix files and the ancillary response files, respectively 
\citep{Ishisaki2007}.
For the following spectral analysis, we used the XSPEC software, version 12.7.0 
\citep{Arnaud1996}. 
The data of the FI and BI CCDs were simultaneously analyzed, but only the FI spectra 
are shown throughout this paper for simplicity. 

\subsection{Galactic X-ray Background}
\begin{figure}[h]
\begin{center}
\includegraphics[width=55mm,angle=-90]{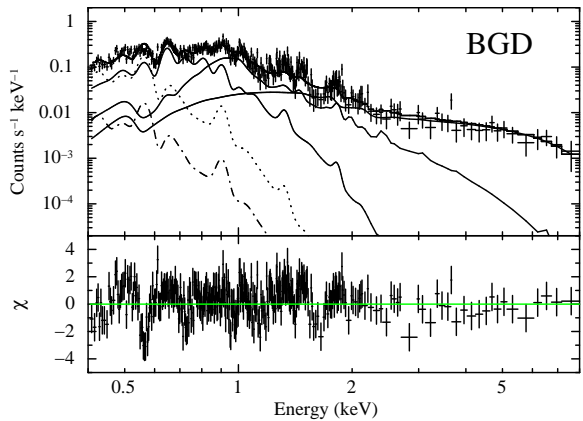}
\caption{Background spectrum fitted with several model components. 
The dashed-dotted and dotted lines show emissions from SB and MWH, respectively. 
The spectra of the local excess and the Lupus Loop are fitted with a power-law and 
two thermal components, respectively (the solid lines).}
\label{fig:bgdfit}
\end{center}
\end{figure}

\begin{table*}[th]
 \caption{Best-fit parameters of the BGD spectrum.}\label{tab:bgd}
  \begin{center}
    \begin{tabular}{lccccc}
       \tableline 
      \tableline
         Component &  $N\rm _H$ ($\times10^{20}$\,cm$^{-2}$) & $kT_{\rm e}$ (keV) & $\Gamma$ & flux\tablenotemark{a} ($\times10^{-12}$\,ergs\,cm$^{-2}$\,s$^{-1}$) \\
      \tableline
SB (CIE)\tablenotemark{b}      & \nodata & 0.1 (fixed) & \nodata & $<0.55$\\
MWH (CIE)\tablenotemark{b}      & 5.6 (fixed) & 0.1 (fixed) & \nodata & $2.43\pm0.21$\\
Local Excess (Power-law)      & 1.0 (fixed) & \nodata & $1.9\pm0.1$ & $1.68\pm0.14$\\
The Lupus Loop1 (CIE)\tablenotemark{c}      & 1.0 (fixed) & $0.22\pm0.01$ & \nodata & $3.22\pm0.14$\\
The Lupus Loop2 (CIE)\tablenotemark{c}      & 1.0 (fixed) & $0.98\pm0.01$ & \nodata & $1.28\pm0.07$\\
                             \tableline
  $\chi ^2$/dof &&&& $439/355=1.24$\\
\tableline
\end{tabular}
    \tablenotetext{a}{Flux in the 0.2--10.0\,keV band.}
    \tablenotetext{b}{Abundances were fixed to 1\,solar.}
    \tablenotetext{c}{Abundances were fixed to 0.2\,solar.}
 \end{center}
\end{table*}

The combined XIS image of SN\,1006 after subtraction of the non-X-ray background (NXB) 
\citep[\textit{xisnxbgen};][]{Tawa2008} is shown in Figure~\ref{fig:fov}.
First, we divided the entire SNR into ``non-thermal'' and ``thermal'' regions and 
extracted spectra using all five of the pointing observations. 
The non-thermal regions consist of the NE and SW rims confined with the dotted lines in Figure 1, while
the thermal region is the whole remaining area.
We subtracted the cosmic X-ray background (CXB) from each spectrum by applying 
a power-law model with a photon index of 1.4 \citep{Kushino2002}. 

The left panel of Figure~\ref{fig:thermal} shows the NXB/CXB-subtracted spectra of 
the non-thermal and thermal regions. 
K-shell lines of O, Ne, Mg, Si, S, Ar, Ca, and Fe were clearly detected in the thermal 
spectrum. To see more detail around the Fe K-shell line, we show the 5--10 keV band 
spectrum in the right panel of Figure~\ref{fig:thermal}. Besides the Fe K-shell line, we 
see a line-like structure at $\sim$5.4\,keV. 

We obtained the Galactic X-ray background (BGD) data from  three near-sky pointing observations of SN\,1006 (BGD: Table~\ref{tab:obs}).
The combined BGD spectrum after the subtraction of the NXB and CXB is shown in 
Figure~\ref{fig:thermal}.  Although the BGD regions (and SN\,1006) are located far off from the Galactic plane, we found clearly noticeable excess X-rays.
The excess X-rays were previously found with the \textit{Tenma} satellite \citep{Koyama1987}. 
Although their origin (thermal or non-thermal) was 
unclear, it was well described by a power-law model with a photon 
index of $2.1\pm0.1$. 
\citet{Ozaki1994} confirmed the excess X-rays with the \textit{Ginga} satellite and 
represented the spectrum using a thermal bremsstrahlung model with a temperature of  7\,keV.  
Soft X-rays from an evolved SNR, the Lupus Loop \citep{Winkler1979}, are widely spread over the full
area of SN\,1006, and hence would contaminate the soft X-ray band.
In addition, the possible contribution of the soft background (SB), which was often 
observed  from  off-plane sky regions such as the North Polar Spur 
\citep{Miller2008} and the Milky Way halo (MWH), should not be ignored.  

Because the observation dates of the SNR Center (AO2) and the BGD regions (PV phase) 
are largely separated, the difference in the energy resolution and event-detection 
efficiency among these observations cannot be ignored. We therefore made a model of the BGD spectrum, and
added it to the source spectrum for the following fitting
procedures, instead of applying direct BGD subtraction.

As we noted, the BGD model should be composed of a power-law plus several thermal 
components with different electron temperatures.  We fit the BGD spectrum with a 
model of these components.
The best-fit results  are shown in Figure~\ref{fig:bgdfit} and 
Table~\ref{tab:bgd}. The photon index of the local excess was obtained to be 
$1.9\pm0.1$, consistent with the result of \citet{Koyama1987}.
We also found that the Lupus Loop has two temperature spectra with $\sim$0.22\,keV 
and $\sim$0.98\,keV.

We then renormalized the flux of the BGD model by the effective area and added it to 
the source spectrum as the background components.
We fixed all the BGD parameters determined in this way.
To confirm the reliability of the background model, we fitted the spectrum of the 
non-thermal rims (NE and SW) with a power-law plus the BGD model, and obtained 
the best-fit photon index of $\sim$2.8. 
This value is consistent with a typical index of the
synchrotron radiation in the X-ray band from the non-thermal
rims of SN\,1006 \citep[e.g.,][]{Koyama1995}.

\subsection{Spectrum of the Entire Thermal Region}\label{sec:fit}
\citet{Yamaguchi2008} found that the Si and S-K$\alpha$ lines in the spectrum of the  
SE region are significantly broadened compared to those expected from a single 
non-equilibrium ionization (NEI) plasma model. 
This broadening was interpreted as a superposition of multiple ejecta components with 
different ionization timescales. 
We found similar line broadening in the entire thermal spectrum, and hence 
applied two NEI plasma components with variable abundances 
\citep[VNEI, NEIvers2.0$+$\footnote{NEI based on improved atomic data including 
inner shell processes. See also \citet{Badenes2006} for more 
detail.}; ][]{Borkowski2001} to represent high- and low-ionization ejecta of 
SN\,1006 (hereafter Ejecta1 and Ejecta2, respectively).
Free parameters were the electron temperature $kT_{\rm e}$, ionization timescale $n_{\rm 
e}t$, emission measure $EM=\int n_{\rm e}n_{\rm H}dV$, and column density 
$N{\rm_H}$ for interstellar absorption. 
Here $n_{\rm e}$, $n_{\rm H}$, $t$, and $V$  are the number densities of electrons 
and protons, elapsed time and the X-ray emitting volume, respectively.

\begin{table*}[!t]
 \caption{Best fit parameters (see Figure~\ref{fig:spec})}\label{tab:para}
  \begin{center}
    \begin{tabular}{llcc}
       \tableline \tableline
          Component & \multicolumn{2}{c}{Parameter} & Value\\
      \tableline
      Absorption & $N\rm _H$ ($\times10^{20}$\,cm$^{-2}$) & & $6.80\pm0.07$\\
      Ejecta1 (VNEI) & $kT_{\rm e}$ (keV) & & $0.48\pm0.01$ \\
                             &  Abundance ($10^4$\,solar) & C & 0 (fixed)\\
                              &                                             & N & 0 (fixed)\\
                              &                                             & O & 1.0 (fixed)\\
                              &                                             & Ne & $0.53\pm0.02$\\
                              &                                             & Mg & $3.71\pm0.07$\\
                              &                                             & Si   & $13.7\pm0.3$\\
                              &                                             & S    & $30\pm1$\\
                              &                                             & Ar   & $27\pm1$\\
                              &                                             & Ca  & $160\pm8$\\
                              &                                             & Fe & $0.56\pm0.01$\\
                              &                                             & Ni  & (=Fe)\\
                              &  $n_{\rm e}t$ (cm$^{-3}$\,s)& & $4.39\pm0.05\times10^{10}$\\
                              &  flux\tablenotemark{a} (ergs\,cm$^{-2}$\,s$^{-1}$) & & $7.85\pm0.01\times10^{-11}$\\
     Ejecta2 (VNEI) & $kT_{\rm e}$ (keV) & & $1.73\pm0.03$\\
                              &  Abundance ($10^4$\,solar) & C & 0 (fixed) \\
                              &                                             & N & 0 (fixed)\\
                              &                                             & O & 1.0 (fixed)\\
                              &                                             & Ne & $0.53\pm0.01$\\
                              &                                             & Mg & $2.3\pm0.1$\\
                              &                                             & Si   & $23.9\pm0.4$\\
                              &                                             & S    & $25.2\pm0.8$\\
                              &                                             & Ar   & $26\pm3$\\
                              &                                             & Ca  & (=Fe)\\
                              &                                             & Fe  & $43\pm3$\\
                              &                                             & Ni  & (=Fe)\\             
                              &  $n_{\rm e}t$ (cm$^{-3}$\,s)& & $1.40\pm0.01\times10^{9}$\\
                              &  flux\tablenotemark{a} (ergs\,cm$^{-2}$\,s$^{-1}$) & & $8.66\pm0.01\times10^{-11}$\\
     ISM (NEI)        & $kT_{\rm e}$ (keV) & & $0.4\pm0.1$\\
                            & $n_{\rm e}t$ (cm$^{-3}$\,s)& & $5.1^{+0.8}_{-0.6}\times10^{9}$\\
                              &  flux\tablenotemark{a} (ergs\,cm$^{-2}$\,s$^{-1}$) & & $4.60\pm0.01\times10^{-11}$\\
     Power-law       & $\Gamma$ & & $3.11\pm0.02$\\
                            &  flux\tablenotemark{a} (photons\,cm$^{-2}$\,s$^{-1}$) & & $3.64\pm0.03\times10^{-2}$\\
                             \tableline
      Emission Line              & Center Energy (keV) &  Normalization (photons\,cm$^{-2}$\,s$^{-1}$) & 1-$\sigma$ Width (eV)\\
      \tableline
                              Fe-L + \ion{O}{7}-K& $0.73\pm0.01$ & $3.92\pm0.08\times10^{-3}$ & 0 (fixed)\\
                               Ca-K & 3.69 (fixed) &  $1.8\pm0.4\times10^{-5}$ & 0 (fixed)\\
                              Cr-K &  $5.42^{+0.20}_{-0.11}$ & $3\pm2\times10^{-6}$ & 0 (fixed)\\ 
                              Fe-K &  $6.45\pm0.02$ & $2.6\pm0.4\times10^{-5}$  & $101^{+33}_{-27}$\\ 
                          \tableline
$\chi ^2$/dof & & & $1225/838=1.46$\\
      \tableline
    \end{tabular}
    \tablenotetext{a}{Flux in the 0.2--10.0\,keV band.}
 \end{center}
\end{table*}

\begin{figure}[t]
\begin{center}
\includegraphics[width=55mm,angle=-90]{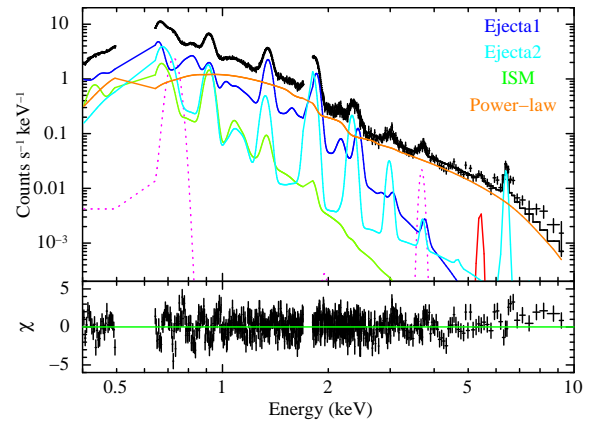}
\caption{FI spectrum of the thermal region fitted with the four-component model plus additional Gaussian lines: Ejecta1 (blue), Ejecta2 (light blue), ISM (light green), a non-thermal power-law (orange), and the magenta and red Gaussian lines for missing lines (see text). 
}
\label{fig:spec}
\end{center}
\end{figure}

Because Type Ia SN ejecta should have a pure-metal
composition with little contribution from H and He, we fixed
the abundance of oxygen \citep{Anders1989} to be a sufficiently high
value ($1\times10^4$\,solar), so that the bremsstrahlung from these elements
is negligible compared to those originating from the heavy element ions. 
Abundances of C and N were fixed at 0, while those of Ne, Mg, Si, S, Ar, and Fe were 
allowed to vary freely and that of Ni was linked to Fe.
For Ejecta2, the Ca abundance was also linked to Fe, but for Ejecta1, Ca abundance was a free parameter because 
L-shell lines of Ca may contribute to the spectrum in the soft X-ray band ($<0.5$\,keV; 
see  Ejecta-1 component in Figure~\ref{fig:spec}). 

Because the ionization time scale for Fe is very low \citep{Yamaguchi2008}, Ca would 
also emit K-shell lines from low ionization states (Ejecta2). 
However, the current NEI model does not include such emission lines.  We therefore added 
a Gaussian line at 3.69\,keV to represent the low-ionization Ca K-shell line
in Ejecta2.
We also added an NEI model with the solar abundances and a power-law 
to represent the swept-up ISM and the non-thermal emissions, respectively. 
The energy band around the neutral Si K edge (1.7--1.8\,keV) was ignored 
because the current response function was not accurate in this limited energy 
band.\footnote{http://heasarc.nasa.gov/docs/suzaku/analysis/sical.html}
We also eliminated the 0.5--0.63\,keV band because the calibration of the 
contamination on the optical blocking filter was
problematic.\footnote{http://www.astro.isas.jaxa.jp/suzaku/doc/suzaku\_td/} 

In this initial fit, we found  a ``shoulder''-shaped residual at around 0.7--0.8\,keV. 
This feature was already noticed by \citet{Yamaguchi2008} and interpreted to be 
a higher transition series of the \ion{O}{7} K-shell lines 
(i.e., K$\delta$ and K$\epsilon$, which are not incorporated in public NEI models). 
Because our NEI model also lacks the \ion{O}{7} K-shell lines higher than K$\delta$, 
we added two Gaussian lines at 723\,eV and 730\,eV to represent K$\epsilon$ and K$\zeta$,
 respectively.
The intensity ratio of K$\zeta$/K$\epsilon$ was assumed to be 0.5 
\citep{Yamaguchi2008} or 0.75 \citep{Broersen2012}.
Although this fit was significantly improved ($\chi ^2$/dof was reduced from 1300/839 
to 1225/838) in both cases, we need an unreasonably high intensity ratio of 
K$\epsilon$ and K$\zeta$ compared to that of lower-excitation level 
(e.g., K$\epsilon$/K$\delta \sim3.4$). 
This may require an additional line for the excess around 0.7\,keV. 

Usually, this energy band is dominated by the \ion{Fe}{17} L-shell emission of the 
3s$\rightarrow$2p transition with  flux comparable to the emission of the 3d$\rightarrow$2p 
transition at $\sim0.8$\,keV \citep{Foster2012}. 
However, in an extremely low-ionized plasma, L-shell
emissions from even lower-charged Fe are also expected to
occur following inner L-shell ionization. 
Although they can contribute to the spectrum around 0.7\,keV, 
the current NEI code does not include any emission from these ions.
In fact, 
our initial best-fit $kT_{\rm e}\sim1.7$\,keV and $n_{\rm e}t\sim1\times10^{9}$\,cm$^{-3}$\,s (for 
Ejecta2) predicts that the most dominant charge state of Fe is less than Fe$^{16+}$. 
Another young Type Ia SNR, E0509$-$67.5, which also has an extremely low-ionization 
timescale of Fe \citep[$n_{\rm e}t<2\times10^{9}$\,cm$^{-3}$; ][]{Kosenko2008}, 
shows a similar excess around 0.7\,keV \citep{Warren2004}. 
The weakness of the \ion{O}{7} and \ion{O}{8} emissions in SNR E0509$-$67.5 
\citep{Warren2004, Badenes2008} leads the excess to be mainly owing to the Fe L-shell emission. 
Because the \ion{O}{7} K$\alpha$ line is strong  in SN\,1006 (Figure~\ref{fig:spec} 
and Table~\ref{tab:para}), both the higher K-shell transitions of \ion{O}{7} and Fe L-shell 
emissions should be considered in the excess around 0.7\,keV.  
For simplicity, we combined all the relevant lines (\ion{O}{7}-K$\epsilon$, K$\zeta$, and Fe L 3d$\rightarrow$2p transitions)  to a 
single Gaussian line at around 0.7\,keV.
As we noted in the previous subsection, we found a line-like structure at $\sim5.4$\,keV.  We therefore added a Gaussian line in the final fitting. Then the center energy and flux were determined to be
$5.42^{+0.20}_{-0.11}$\,keV and  $3{\pm2}$\,photons\,cm$^{-2}$\,s$^{-1}$, respectively.

\begin{figure*}[t]
\begin{center}
\includegraphics[width=170mm]{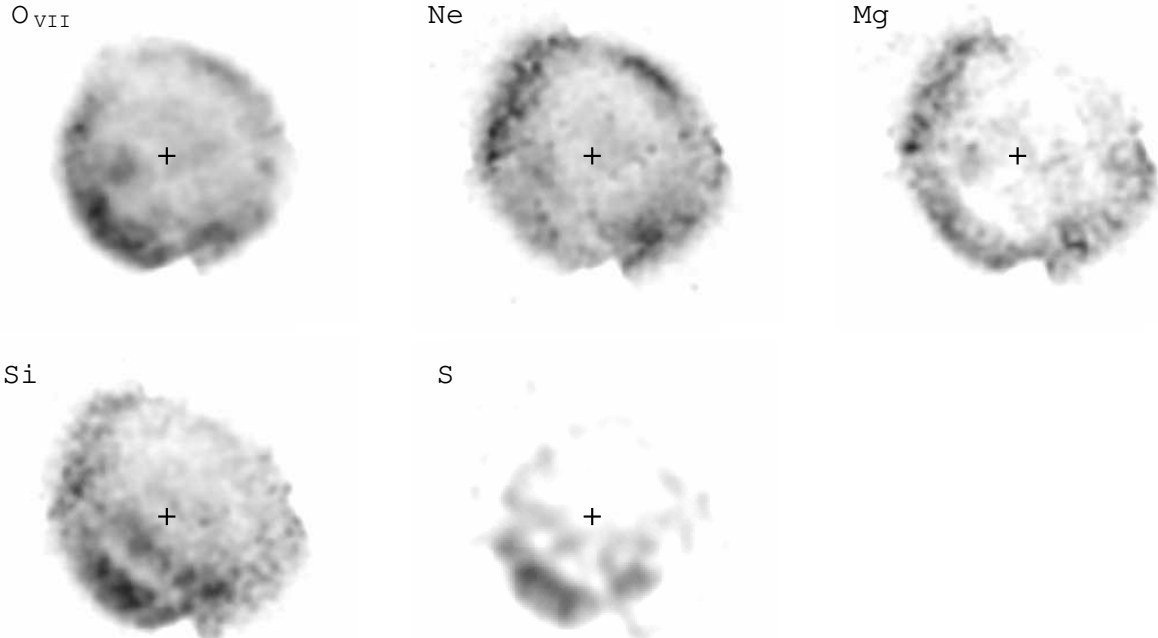}
\caption{Vignetting-corrected intensity maps of the K$\alpha$ lines from O, Mg, and Si. The geometric center of SN\,1006 is indicated with the cross marks.}
\label{fig:band}
\end{center}
\end{figure*}

The best-fit results of this final model are given in Figure~\ref{fig:spec} and 
Table~\ref{tab:para}. 
The results are basically the same as those of \citet{Yamaguchi2008}; the spectrum 
consists of a power-law, and plasmas of ISM and two ejecta components 
with  different ionization timescales.
The abundances of  both  ejecta 
components are also consistent with each other for Ne through Ar within a factor of $\sim$2, 
whereas the Fe abundance in Ejecta2 is more than  one order of magnitude larger than 
that in Ejecta1.

In Figure~\ref{fig:spec}, we see excess structures at both sides of the line at 
$\sim 6.4$~keV.  This feature is likely owing to the broadening of the Fe K$\alpha$ line. 
We therefore fit the line with a Gaussian model and found the line energy and width  to be $6.45{\pm0.02}$\,keV and $101^{+33}_{-27}$\,eV, respectively 
(Table~\ref{tab:para}). 
Although Ejecta1 has an extremely large abundance of Ca (Table~\ref{tab:para}),
this value should be considered cautiously because  the Ca abundance is mainly 
determined by the L-shell lines for which uncertainty in the atomic data is large.

We found that the contribution of the ISM component is relatively lower than the result 
of SE \citep[Figure~8 in][]{Yamaguchi2008},  but is consistent with the 
interpretation by \citet{Cassam2008}. The O emission in the entire SNR predominantly 
originates from the shock heated ejecta rather than from the ISM. 

\begin{figure*}[t]
\begin{center}
\includegraphics[width=120mm]{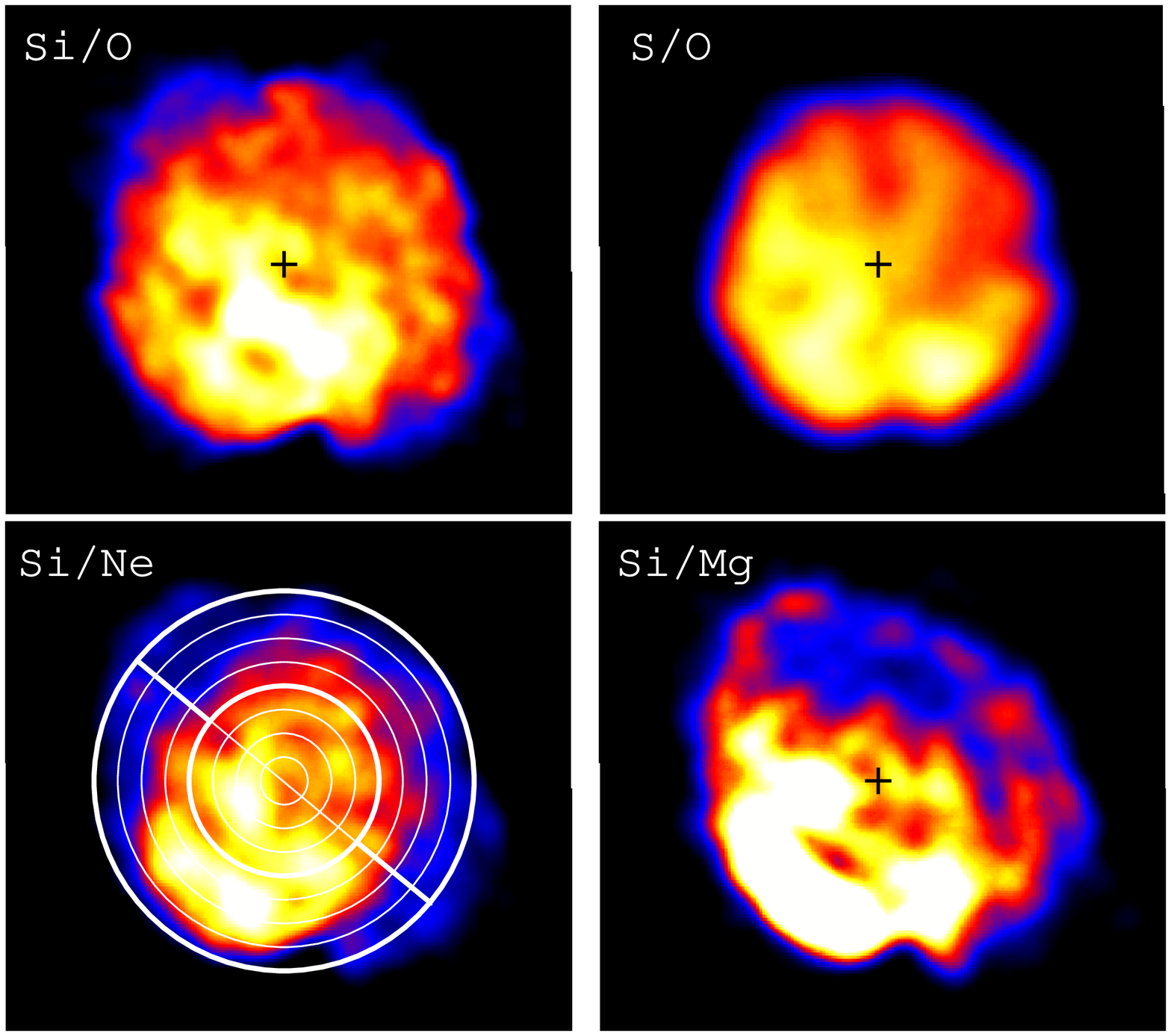}
\caption{Images of the line flux ratio of the K$\alpha$ lines from Si to O, S to O, Si 
to Ne, and Si to Mg. The geometric center of SN\,1006 is shown with cross marks. 
The spectral extraction regions are shown with  thin white lines. The width of 
each annulus is 2\arcmin. Thick white lines divide the remnant into three parts 
named as  S (southeast), C (center), and N (northwest).}
\label{fig:ratio}
\end{center}
\end{figure*}

\subsection{Spatially Resolved Analysis}
Figure~\ref{fig:band} shows the vignetting-corrected narrow band images of the 
K$\alpha$ lines from O, Ne, Mg, Si, and S, after subtraction of the underlying 
continuum levels estimated by the interpolation of  the adjacent band fluxes. 
The images show rim-brightening morphology, particularly in the light elements such 
as O, Ne, and Mg.  In the interior regions, O, Ne, and Mg are rather  uniformly
distributed.  In contrast, the distributions of Si and S are more asymmetric in the 
inner region; the SE rim is much brighter than the other regions, and the ``second 
shell'' can be seen at the middle between the outermost shell and the SNR's center. 
We also present the ratios of Si/O, S/O, Si/Ne, and Si/Mg in 
Figure~\ref{fig:ratio}.  All the ratios show  clear increases toward the SE 
rim, suggesting asymmetric ejecta distribution for the heavier elements (i.e., Si, 
S).  

\begin{figure*}[t]
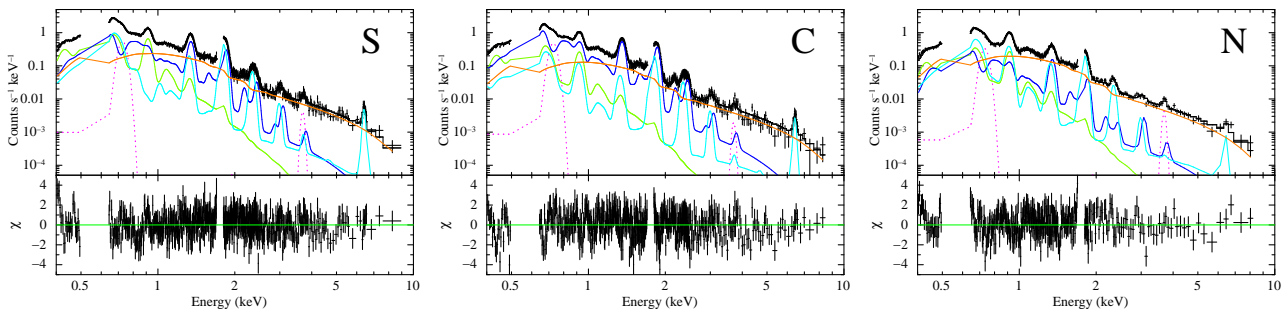

\begin{center}
\includegraphics[width=40mm,angle=-90]{figure7a.eps}
\includegraphics[width=40mm,angle=-90]{figure7b.eps}
\includegraphics[width=40mm,angle=-90]{figure7c.eps}
\caption{FI spectrum of SN\,1006 obtained from the regions S, C, and N. Each spectrum 
was fitted with the same four-component model as the whole region.}
\label{fig:spec_scn}
\end{center}
\end{figure*}

To investigate more quantitatively, we divided the entire
SNR into three regions: a center circle of 8\arcmin-radius (C), and half-annuluses for the northwest (N) and southeast (S)
(the thick white lines in Figure~\ref{fig:ratio}).
The geometric center of the remnant was defined to be ($\alpha$, $\delta$)$_{\rm 
J2000.0}$ = (225.7371, --41.9336). 
The spectra of these three regions are shown in Figure~\ref{fig:spec_scn}.
We fitted the spectra with the same four-component model (and the same assumptions) 
applied for the entire thermal spectrum.
The best-fit results are given in Table~\ref{tab:para_scn}. 
We found that the heavier elements are indeed more abundant in the region S  
than in the region N. 

\begin{figure*}[th]
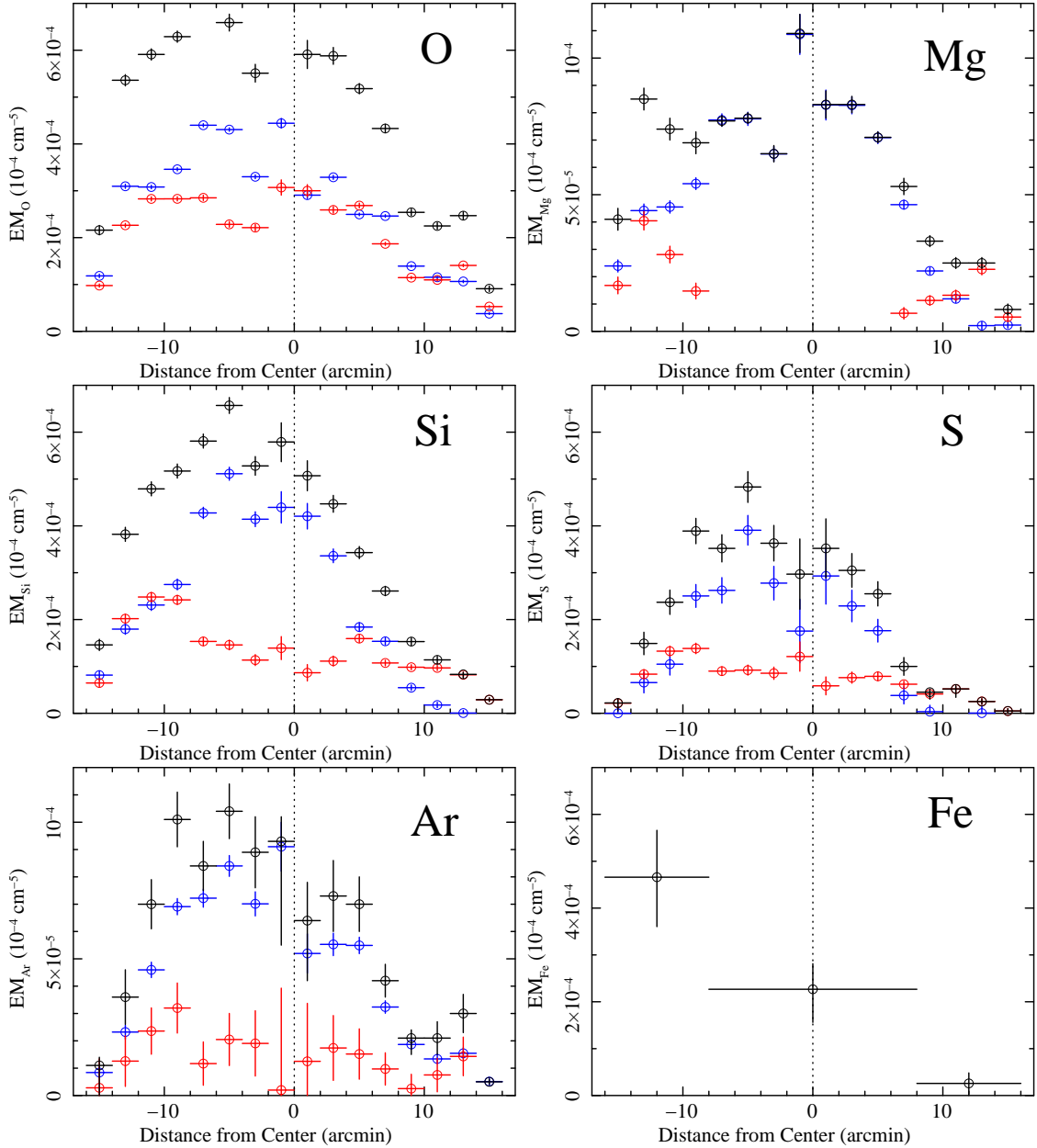

\begin{center}
\includegraphics[width=55mm,angle=-90]{figure8a.eps}
\includegraphics[width=55mm,angle=-90]{figure8b.eps}
\includegraphics[width=55mm,angle=-90]{figure8c.eps} 
\includegraphics[width=55mm,angle=-90]{figure8d.eps}
\includegraphics[width=55mm,angle=-90]{figure8e.eps}
\includegraphics[width=55mm,angle=-90]{figure8f.eps}
\caption{Radial profiles of the emission measure ($EM$) for various elements (O, Mg, Si, S, Ar, and Fe) in  ejecta. Red, blue, and black represent the results of Ejecta1, Ejecta2 (Table~\ref{tab:para}), and summation of them, respectively. Note that the vertical scales of the $EM$ of Mg and Ar ($EM_{\rm{Mg, Ar}}$) are 1/5 of those of O, Si, S, and Fe ($EM_{\rm{O, Si, S, Fe}}$). The horizontal axis shows the distance from the geometric center of SN\,1006; the positive and negative values are the NW and  SE sides, respectively.}
\label{fig:em}
\end{center}
\end{figure*}

We further divided the whole SNR  into 16 regions: 8 annuli with widths of 2\arcmin , 
each divided into  half-rings of the northwest (NW) and southeast (SE) areas (solid lines 
in Figure~\ref{fig:ratio}). Then we fitted each spectra with the same four-component 
model. 
Because the statistics of the Fe line were insufficient, we fixed its abundance to 
the best-fit value in Table~\ref{tab:para_scn}.
The best-fit $EM$s for each element in  ejecta (Electa1+ Ejecta2) are 
given in Figure~\ref{fig:em}. 

\begin{figure}[ht]
\begin{center}
\includegraphics[width=60mm,angle=-90]{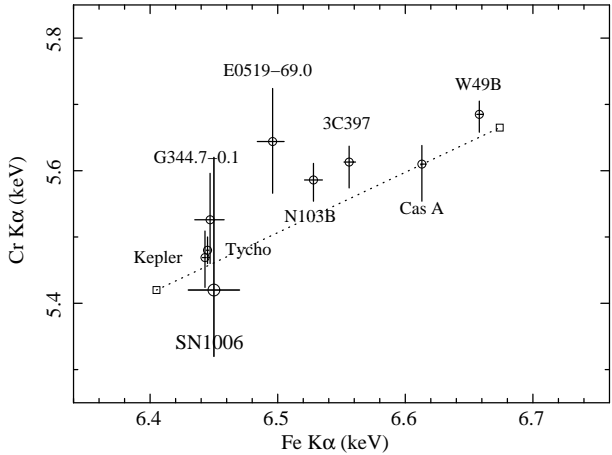}
\caption{The K-shell line  energy of Fe versus that of Cr in SN\,1006.  We also plotted previous results of W49B \citep{Hwang2000}, Tycho \citep{Tamagawa2009}, 
G\,344.7-0.1 
\citep{Yamaguchi2012}, and other remnants \citep{Yang2013} with circles. Open 
squares at the lower-left and upper-right represent the values for the neutral and 
He-like states, respectively. }
\label{fig:CrFe}
\end{center}
\end{figure}

\begin{table*}[ht]
 \caption{Best fit parameters (Figure~\ref{fig:spec_scn}).}\label{tab:para_scn}
  \begin{center}
    \begin{tabular}{lllccc}
       \tableline \tableline
          Component & \multicolumn{2}{c}{Parameter} & \multicolumn{3}{c}{Value}\\
          & & & Region S & Region C & Region N\\
      \tableline
      Absorption & $N\rm _H$ ($\times10^{20}$\,cm$^{-2}$) & 		& $6.8\pm0.1$     				& $6.8\pm0.1$ 					& $6.8\pm0.2$\\
      Ejecta1 (VNEI) & $kT_{\rm e}$ (keV) &                                      		& $0.95\pm0.01$				& $0.95\pm0.01$ 				& $1.02^{+0.04}_{-0.03}$\\
                             &  Abundance ($10^4$\,solar) 	& C  		& 0 (fixed)	           				& 0 (fixed)		       	       			& 0 (fixed)	\\
                              &                                             		& N 		& 0 (fixed)			 			& 0 (fixed)				 		& 0 (fixed)	\\
                              &                                             		& O 		& 1.0 (fixed) 					& 1.0 (fixed) 					& 1.0 (fixed)\\
                              &                                             		& Ne		& $0.45\pm0.03$ 				& $0.54\pm0.02$ 				& $0.45\pm0.07$\\
                              &                                             		& Mg	& $3.0\pm0.1$ 					& $2.17\pm0.05$ 				& $2.4\pm0.2$\\
                              &                                             		& Si   	& $6.8\pm0.2$ 					& $6.7\pm0.2$ 					& $2.7\pm0.4$\\
                              &                                             		& S    	& $7.3\pm0.8$ 					& $8.8\pm0.7$ 					& $<2$\\
                              &                                             		& Ar   	& $28\pm2$ 					& $14\pm1$ 					& $35\pm5$\\
                              &                                             		& Ca  	& $28\pm2$ 					& $34\pm12$ 					& $110\pm50$\\
                              &                                             		& Fe 	& $0.65\pm0.02$ 				& $0.40\pm0.01$ 				& $0.46\pm0.05$\\
                              &                                             		& Ni  	& (=Fe) 						& (=Fe) 						& (=Fe)\\
                              &  $n_{\rm e}t$ (cm$^{-3}$\,s)& 					& $2.66\pm0.09\times10^{10}$ 	& $1.91\pm0.05\times10^{10}$ 	& $1.8\pm0.2\times10^{10}$\\
            &  flux\tablenotemark{a} (ergs\,cm$^{-2}$\,s$^{-1}$) & 	& $1.02\pm0.01\times10^{-11}$ 	& $1.73\pm0.01\times10^{-11}$ 	& $0.32\pm0.01\times10^{-11}$\\
     Ejecta2 (VNEI) & $kT_{\rm e}$ (keV) & 						& $1.83^{+0.05}_{-0.04}$ 		& $2.50\pm0.09$ 				& $2.74^{+0.18}_{-0.14}$\\
                              &  Abundance ($10^4$\,solar) 	& C 		& 0 (fixed)						& 0 (fixed) 					& 0 (fixed)	\\
                              &                                             		& N 		& 0 (fixed) 					& 0 (fixed)	 					& 0 (fixed)	\\
                              &                                            		& O 		& 1.0 (fixed) 					& 1.0 (fixed) 					& 1.0 (fixed)\\
                              &                                             		& Ne 	& $<0.02$		 				& $0.41\pm0.06$ 				& $0.58\pm0.03$\\
                              &                                             		& Mg 	& $2.1\pm0.2$ 					& $3.3\pm0.6$ 					& $2.5\pm0.2$\\
                              &                                             		& Si   	& $24\pm1$ 					& $37\pm2$ 					& $12.9\pm0.7$\\
                              &                                             		& S    	& $30\pm2$ 					& $34\pm3$ 					& $10\pm1$\\
                              &                                             		& Ar   	& $14\pm7$ 					& $32\pm10$ 					& $<5$\\
                              &                                             		& Ca  	& (=Fe) 						& (=Fe) 						& (=Fe)\\
                              &                                             		& Fe  	& $20\pm5$ 					& $17\pm5$ 					& $3\pm2$\\
                              &                                             		& Ni  	& (=Fe) 						& (=Fe) 						& (=Fe)\\             
                              &  $n_{\rm e}t$ (cm$^{-3}$\,s)& 					& $1.58\pm0.02\times10^{9}$ 		& $1.31\pm0.04\times10^{9}$ 		& $1.41\pm0.02\times10^{9}$\\
                  &  flux\tablenotemark{a} (ergs\,cm$^{-2}$\,s$^{-1}$) & 	& $2.08\pm0.04\times10^{-11}$	& $1.29\pm0.01\times10^{-11}$ 	& $1.47\pm0.01\times10^{-11}$\\
     ISM (NEI)        & $kT_{\rm e}$ (keV) & 						& 0.4 (fixed) 					& 0.4 (fixed)					& 0.4 (fixed)\\
                            & $n_{\rm e}t$ (cm$^{-3}$\,s)& 					& $5.1\times10^{9}$ (fixed) 		& $5.1\times10^{9}$ (fixed)  		& $5.1\times10^{9}$ (fixed)\\
                  &  flux\tablenotemark{a} (ergs\,cm$^{-2}$\,s$^{-1}$) & 	& $2.22\pm0.01\times10^{-11}$	& $1.76\pm0.01\times10^{-11}$ 	& $0.91\pm0.01\times10^{-11}$\\
     Power-law       & $\Gamma$ & 							& $2.98\pm0.03$ 				& $3.16\pm0.04$ 				& $3.27\pm0.03$\\
             &  flux\tablenotemark{a} (photons\,cm$^{-2}$\,s$^{-1}$) & 	& $0.68\pm0.01\times10^{-2}$ 	& $0.64\pm0.01\times10^{-2}$ 	& $0.63\pm0.01\times10^{-2}$\\
 \tableline
                              \tableline
$\chi ^2$/dof & & 										& $670/604=1.11$ 				& $751/569=1.32$ 				& $560/511=1.10$\\

      \tableline
    \end{tabular}
    \tablenotetext{a}{Flux calculated 0.2\,keV to 10.0\,keV.}
 \end{center}
\end{table*}

\section{Discussion}\label{sec:discussion}
\subsection{Elemental Abundances in  Ejecta}
In section~\ref{sec:fit}, we successfully modeled the spectrum of the entire thermal 
region with the components applied previously by \citet{Yamaguchi2008}. 
The emission lines of the heavy elements exhibit clear evidence for 
over-abundance relative to the solar values, indicating the SN ejecta origin. 
The difference in the ionization age between the two
components suggests that Ejecta2 was heated more recently
than Ejecta1.
The suppressed relative abundance of Fe in Ejecta1 (Table~\ref{tab:para}) implies the absence of Fe in the outer ejecta layer,
consistent with the previous claim by \citet{Yamaguchi2008}.

We find a possible line feature at $\sim 5.4$\,keV. This centroid
energy suggests its origin to be a Cr K$\alpha$ emission. To examine
this possibility, we plot in Figure~\ref{fig:CrFe} the Cr K$\alpha$ and Fe K$\alpha$
centroids observed in other SNRs \citep[e.g.,][]{Yang2013}
and compare with our results. We find that the values from SN\,1006 are reasonably close to the previous Cr K$\alpha$--Fe K$\alpha$ centroid
correlations, supporting our claim of the first detection of
Cr from this remnant.

\begin{figure}[t]
\begin{center}
\includegraphics[width=75mm]{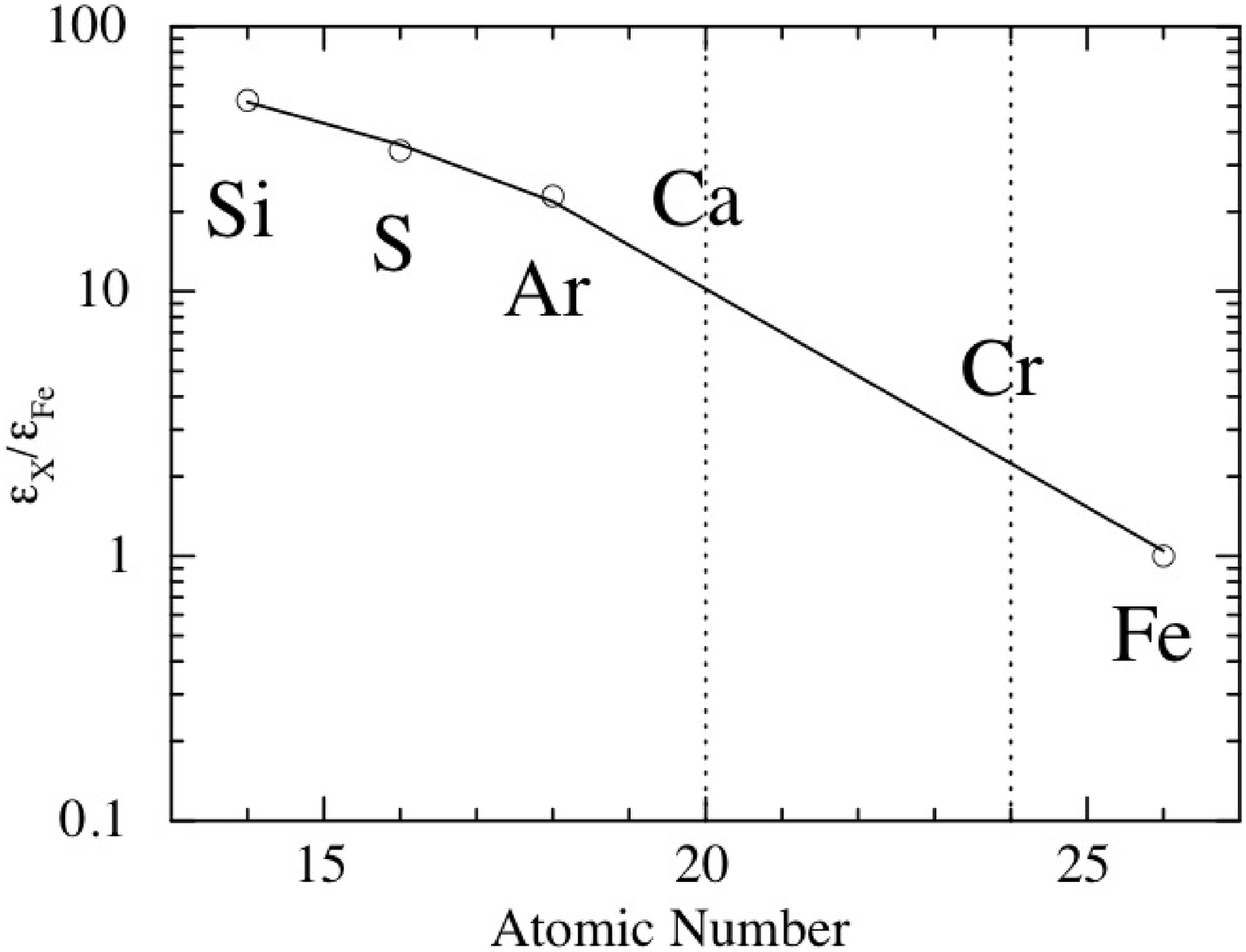}
\caption{Emissivity of heavy elements relative to that of Fe. These values were 
calculated from the NEI model of the best-fit Ejecta2 temperature and ionization parameter. The line is a spline fit to them.}
\label{fig:emissivity}
\end{center}
\end{figure}

Because atomic data for Cr are not available in the current NEI model, 
we estimate the emissivity $\varepsilon_{\rm{Cr}}$ following 
the method of the ASCA measurement for W49B \citep{Hwang2000}. 
Using the best-fit temperature and the ionization timescale for  Ejecta2 
component, we calculate emissivities for Si, S, Ar, and Fe. 
The results are given in Figure~\ref{fig:emissivity}. 
Interpolating these values to that of Cr, we estimated the Cr/Fe abundance 
ratio $Z_{\rm{Cr}}/Z_{\rm{Fe}}$ to be  $2.5\pm1.9$\,solar. Here we assumed that Cr 
is mainly contained in Ejecta2, as is the case for Fe. 
 
As we noted, the NEI model we used also lacks emission data for Ca in the low-ionization 
state (Ejecta2). Therefore, we estimated $Z_{\rm{Ca}}/Z_{\rm{Fe}}$ with the same 
method  to be $0.8\pm0.2$ in Ejecta2. This result is consistent with our initial 
assumption that the abundance of Ca is nearly equal to Fe in Ejecta2.

\begin{figure}[t]
\begin{center}
\includegraphics[width=60mm,angle=-90]{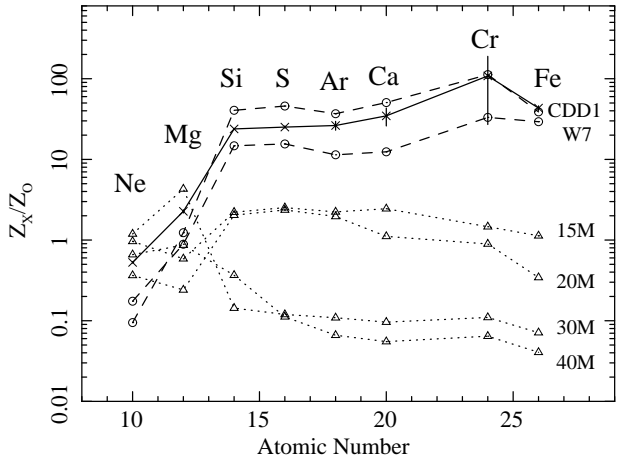}
\caption{Metal abundances of Ejecta2 relative to O as a function of atomic 
number (the solid line and X marks).  The dashed lines and circles  represent the 
Delayed Detonation, i.e., CDD1 (one of the DDT models) and classical deflagration 
model, W7 \citep{Iwamoto1999}. The dotted lines and triangles represent 
core-collapse models with main sequence masses of 
15$M_\odot$, 20$M_\odot$, 30$M_\odot$, and 40$M_\odot$ \citep{Woosley1995}.} 
\label{fig:metal}
\end{center}
\end{figure}

We compare the abundances of Ejecta2 (relative to O) 
with the nucleosynthesis yields predicted by several theoretical models 
(Figure~\ref{fig:metal}).
We employ the Type Ia deflagration or delayed-detonation models 
\citep{Iwamoto1999} and the core-collapse SN models with various progenitor masses
\citep[15$M_\odot$, 20$M_\odot$, 30$M_\odot$, and 40$M_\odot$;][]{Woosley1995}.
The observed abundance pattern is confirmed to be broadly consistent with the standard 
Type Ia models.
The abundance ratios of Si/O, S/O, Ar/O, are Ca/O are enhanced compared to 
those predicted by the W7 deflagration model, and slightly closer to the yields of 
the delayed-detonation explosion (CDD1).

\begin{figure*}[th]
\begin{center}
\includegraphics[width=120mm]{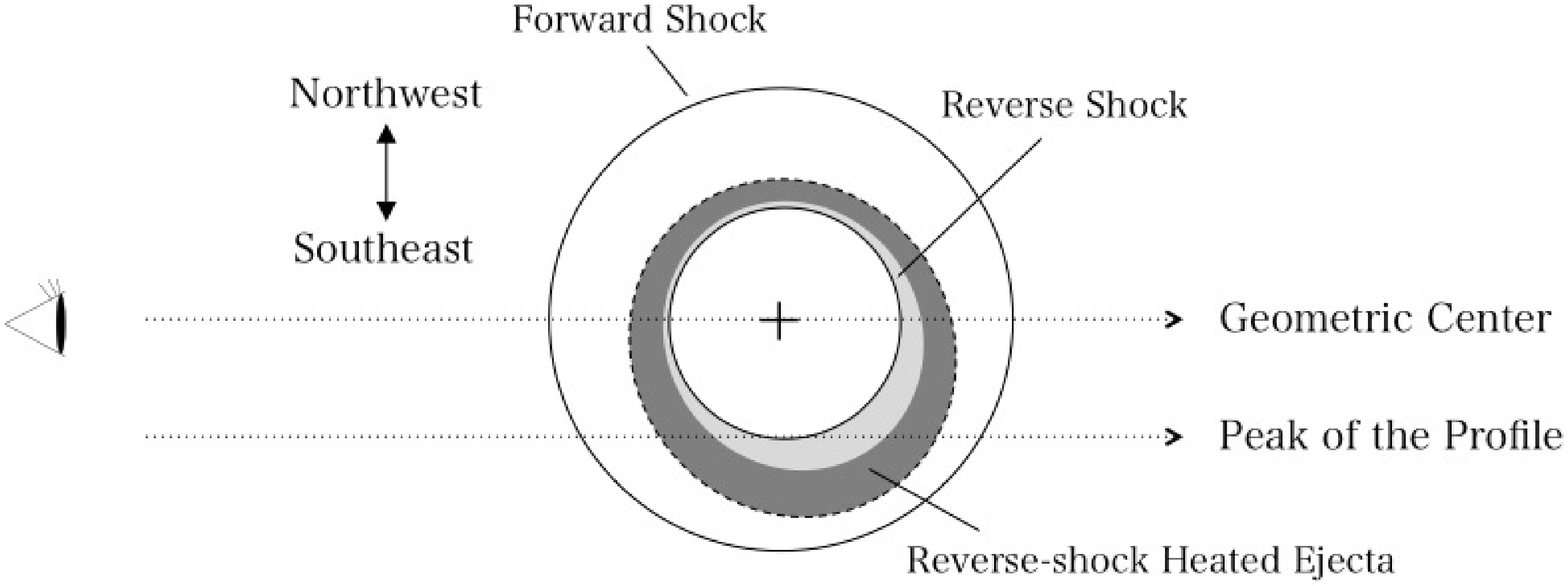}
\caption{Schematic diagram of the ejecta structure of SN\,1006. 
The gray area represents  ejecta already heated by the reverse shock.
The light gray and dark gray regions correspond to the Fe-rich core and the Si-rich 
layer, respectively. 
The heavy elements are displaced to the far-side and the SE (see text).}
\label{fig:diagram}
\end{center}
\end{figure*}

\subsection{Three-dimensional Geometry of SN\,1006}
The spherical outer shell with  rim-brightening in O, Ne, and Mg (Figure~\ref{fig:band})  would be mainly owing to the ISM components, because  ejecta distributions in Figure~\ref{fig:em} show no enhancement in these regions.
On the other hand,  the interior is likely to be dominated by  ejecta (e.g., 
Figure~\ref{fig:em}). Then  ejecta distributions are broadly separated into two 
patterns: the uniform 
interior for the lighter elements (O, Ne, and Mg) and more asymmetric distributions of 
the heavier elements (Si and S) with peak positions offset by 
$5\arcmin$ ($\sim$3.2\,pc).  The off-center distribution of the heavier elements is
also confirmed in Figure~\ref{fig:ratio}. 
Notably, these two groups (O-Ne-Mg and Si-S) are
synthesized in different nuclear burning processes in Type
Ia SNe explosions: C-burning for O-Ne-Mg compared to O-burning
and incomplete Si-burning for Si-S. We thus presume that the
latter products distribute more asymmetrically than the former products.

The asymmetry of  ejecta should be formed either at the time of the SN 
explosion or in the process of the following interaction with the ISM.
If the ambient density in the SE is higher than in the other regions, 
a reverse shock might heat up ejecta earlier, which may explain the relative enhancement of the $EM$ in the SE. 
To estimate the ambient densities  in  the SE and  NW rims, we analyzed spectra from  
an annulus  at the outer edges of the SE and NW.  
We fitted the spectra and found the best-fit  normalizations ($EM$: emission measure)  of the ISM components to be 
$2.46\pm0.29\times10^{-4}$\,cm$^{-5}$ and $1.93\pm0.22\times10^{-4}$\,cm$^{-5}$ for the SE and NW,  respectively.
Given that SN\,1006 is a sphere with a radius of 10.2\,pc, the volume of the annulus  is estimated to be 35\,pc$^3$.
Then, the ISM densities are $n_{\rm H}=0.11\pm0.01$\,cm$^{-3}$ and $0.09\pm0.01$\,cm$^{-3}$.
The ambient densities ($n_{\rm H}/4)$ are thus $\sim0.03$\,cm$^{-3}$ and $\sim0.02$\,cm$^{-3}$,
for  SE and NW, respectively.
Previous observations \citep[e.g.,][]{Heng2007} indicate that the ambient density in the NW 
is 0.15--0.3\,cm$^{-3}$, significantly  higher than our result.
Because the  best-fit normalization ($EM$)  was coupled  with that of $n_{\rm e}t$  and  anti-correlated to 
$kT_{\rm e}$,  a higher value of $n_{\rm H}$/4 (up to $\sim0.1$\,cm$^{-3}$) is statistically acceptable in the NW.
The annulus we used for the spectrum is broader than the outermost edge of the recent shock encounter.
On the other hand, the forward shock in the NW  has only interacted with the denser region fairly recently \citep{Katsuda2013}, which may explain the discrepancy between our result and the previous studies.

Accordingly, within possible statistical and systematic uncertainties,  we can  conclude 
that the ambient density in the SE is lower than 0.1\,cm$^{-3}$ and does not particularly exceed those in the other rims.
Furthermore, if an inhomogeneous ambient density has affected the large-scale variation of  ejecta, the lighter elements (O, Ne, and Mg) should similarly distribute asymmetrically.
As shown in Figure~\ref{fig:ratio}, the actual distributions clearly refute this hypothesis.
Thus, it is more plausible that ejecta asymmetry was caused by an asymmetric SN explosion.

We found the ``second shell'' in Si and S  at the middle between the outermost shell 
and the SNR's center ($\sim6\arcmin$ from the Center).  
The ``second shell'' is most 
likely from  ejecta, because we see a peak of Si and S at this position in 
ejecta abundance (Figure~\ref{fig:band}). 
Because the foreground absorption is negligible in energies above $\sim1.5$\,keV,  
this second-shell structure is not owing to the absorption effect, but it likely  
originates from the ejecta rim, reminiscent of similar double-shell morphology in 
G299.2-2.9 \citep{Park2007} (although this remnant is dominated by swept-up ISM, 
unlike SN\,1006). 

We also found that the  $EM$ of Fe increases from  NW to SE, as 
\citet{Yamaguchi2008} found that the Fe-rich core spreads out to SE.
These results indicate that  ejecta, particularly heavier
elements in more inner regions, expand preferentially toward the SE.
Based on the broad UV absorption lines of Si and Fe of several background objects, 
\citet{Hamilton1997} estimated that the diameter toward the far side of SN\,1006 
is roughly 20\% larger than the rest of the remnant \citep[see Figure~7 
of][]{Hamilton1997}.
\citet{Winkler2005} also confirmed the ejecta structure of Si and Fe by a similar 
analysis that the shell of SN\,1006 is almost spherical \citep[see Figure~8 
of][]{Winkler2005}.
We thus propose that  ejecta expanded not only \textit{transverse to} but also 
\textit{along} the line of sight;  ejecta were displaced toward the SE and 
far side of the center (Figure~\ref{fig:diagram}).
If the reverse shock heated up a part of  ejecta components, then the centrally peaked 
profile is easily understood  (Figure~\ref{fig:diagram}).
In addition, if the reverse shock recently reached to the surface of the Fe-rich core 
(light gray region in Figure~\ref{fig:diagram}), the $EM$ is higher at the 
shifted direction (lower right in Figure~\ref{fig:diagram}).

Including the ``second shell'', the peak positions of the heavy elements (Si, 
S, and Ar) are $\sim5\arcmin=3.2$\,pc away from the geometric center toward the SE 
(Figure~\ref{fig:em}).
Assuming a distance of 2.2\,kpc and the SNR age of 1000\,yr, we estimate that the 
velocity difference (owing to an asymmetric explosion) between the SE and NW 
direction is  $\sim3100$\,km\,s$^{-1}$ in projection.
We found that Fe K$\alpha$ is broadened by $101^{+33}_{-27}$\,eV (1-$\sigma$), for the first time from SN\,1006. 
If this width is owing to the Doppler broadening of  ejecta, as is the case of Tycho 
\citep{Hayato2010}, the expansion velocity in the line of sight should be
$\sim3000$\,km\,s$^{-1}$. This value is consistent with that of  
\citet{Katsuda2013}, in which they  measured the proper motion of the NW rim in X-ray.

\section{Summary}
This paper focused on the thermal spectrum of SN\,1006 based on the deep observation 
($\sim400$\,ks in total) with \textit{Suzaku} XIS.
The results are summarized as follows:
\begin{enumerate}
\item The X-ray spectra are well represented by a model with three NEI thermal plasmas 
(ejecta with different ionization parameters and ISM) plus one power-law component 
(non-thermal emission).
\item Ejecta abundance ratios of Ne, Mg, Si, S, Ar, Ca, Cr, and Fe relative to 
O are all in good agreement with yields predicted by the standard Type Ia theories: the classical carbon-deflagration model and the deflagration to detonation 
transition mechanism.
\item The distribution of the centroids of Si, S, and Ar are significantly shifted 
from the SNR's geometric center by $\sim3.2$\,pc toward the SE, which we argue is likely 
to be associated with an asymmetric explosion of the progenitor. The velocity  (in 
projection) difference of ejecta  is $\sim$3100\,km\,s$^{-1}$ at the distance 
of 2.2\,kpc.  The morphology of O, Ne, and Mg in the interior are more symmetric. 
\item We found possible evidence for the Cr-K shell line and line broadening in the Fe-K shell emission.
\end{enumerate}

\acknowledgments
The authors thank Dr. T. G. Tsuru and Dr. M. Nobukawa for carefully reading our manuscript.
H.U. is supported by Japan Society for the Promotion of Science (JSPS) 
Research Fellowship for Young Scientists. 
K.K. is supported by JSPS KAKENHI, Grant Numbers 23000004 and 24540229.
This work was supported by the Grant-in-Aid for the Global COE Program``The Next Generation of Physics, Spun from Universality and Emergence'' from the Ministry of Education, Culture, Sports, Science and Technology (MEXT) of Japan.

{\it Facilities:} \facility{\textit{Suzaku} (XIS)}.

\bibliographystyle{apj}
\bibliography{bibtex_library_SN1006}

\end{document}